\documentclass[11pt,twoside]{article}

\usepackage{asp2014}

\aspSuppressVolSlug
\resetcounters

\begin{document}

\title{The recalibration of the UVES-POP stellar spectral library}

% Note the position of the comma between the author name and the 
% affiliation number.
% Author names should be separated by commas.
% The final author should be preceded by "and".
% Affiliations should not be repeated across multiple \affil commands. If several
% authors share an affiliation this should be in a single \affil which can then
% be referenced for several author names.
% See ManuscriptInstructions.pdf and ASPmanual2010.pdf 3.1.4 for more details
\author{Svyatoslav~Borisov,$^{1,2}$ Igor~Chilingarian,$^{1,3}$ Eugene~Rubtsov,$^{1,2}$ C\'edric~Ledoux$^4$, Claudio~Melo$^4$ and Kirill~Grishin$^{1,2}$
\affil{$^1$Sternberg Astronomical Institute, Moscow State University, Moscow, Russia;
\email{slavasai95@gmail.com}}
\affil{$^2$Faculty of Physics, Moscow State University, Moscow, Russia}
\affil{$^3$Smithsonian Astrophysical Observatory, Cambridge, MA, USA; \email{igor.chilingarian@cfa.harvard.edu}}
\affil{$^4$European Southern Observatory, Santiago, Chile}}

% This section is for ADS Processing.  There must be one line per author.
\paperauthor{Svyatoslav~Borisov}{slavasai95@gmail.com}{}{Sternberg Astronomical Institute}{}{Moscow}{}{119992}{Russia}

\begin{abstract}
We have re-reduced all spectra from the UVES-POP stellar spectral library
using the version 5.5.7 of the UVES pipeline and an algorithm we
designed, which allows us to remove ripples in regions where echelle orders
are stitched. These ripples are caused by the offset of a flat field with
respect to a science frame and under- or oversubtraction of scattered light. 
We have also developed an approach to merge 6 UVES spectral chunks divided
by gaps in the spectral coverage by using synthetic
stellar atmospheres to predict the flux difference between the segments.  At
the end, we improved the flux calibration quality to 2\%\ or better for
85\%\ of 430 spectra in the library.
\end{abstract}

\section{Motivation} 
High quality models of stellar populations are in great demand now -- they
are used to interpret galaxy and star cluster spectra to determine the age,
metallicity and chemical abundances.  An essential ingredient of stellar
population synthesis is a library of stellar spectra.  This work is aimed at
the recalibration of the UVES-POP library \citep{2003Msngr.114...10B} to improve
flux calibration to the level of 2\%\ or better required to model stellar
populations.  Such level of precision is a non-trivial task for multi-order
echelle spectra.

\section{Introduction: About UVES-POP and its imperfections}
UVES-POP is a stellar spectral library that includes spectra of $\sim$450
stars collected with UVES (Ultraviolet and Visible Echelle Spectrgraph) \citep{2000SPIE.4008..534D} 
at ESO VLT in 2001-2003.  Some important characteristics of UVES-POP stellar spectra:
\begin{itemize}
\item typical spectral resolution: $R = \lambda/\Delta\lambda = 80000$;
\item the majority of spectra have SNR=300-500 per resolution element in the V band;
\item wide coverage in the $T_{\mathrm{eff}}$, [Fe/H], $log(g)$ parameter space.
\end{itemize}

\subsection{The problem of ripples}
Most of the original UVES-POP spectra have noticeable ripples in the regions
where echelle orders are stitched.  They are caused by (i) the offsets
between science and flat field exposures and (ii) under- or oversubtraction
of scattered light.  Shift of flat-fields may have different reasons, but
the most probable are temperature change in the dome of the telescope
(flat-fields are taken in the evening or morning, but science frames at
night) and mechanical flexures at different parallactic angles.

\subsection{Inaccurate merging of spectral chunks}
Merged UVES spectra have two gaps in the wavelength coverage due to the gaps
between UVES CCDs.  The flux difference on the two sides of each gap
because of the spectral shape was neglected in the original UVES-POP
reduction, and the flux level was equated: this caused ''steps'' in flux.

\subsection{Quality control}
The quality control for the original UVES-POP spectra, which were reduced
using an old UVES pipeline sometimes accepted poor quality spectra but
rejected spectra of reasonable quality. 
\articlefigure{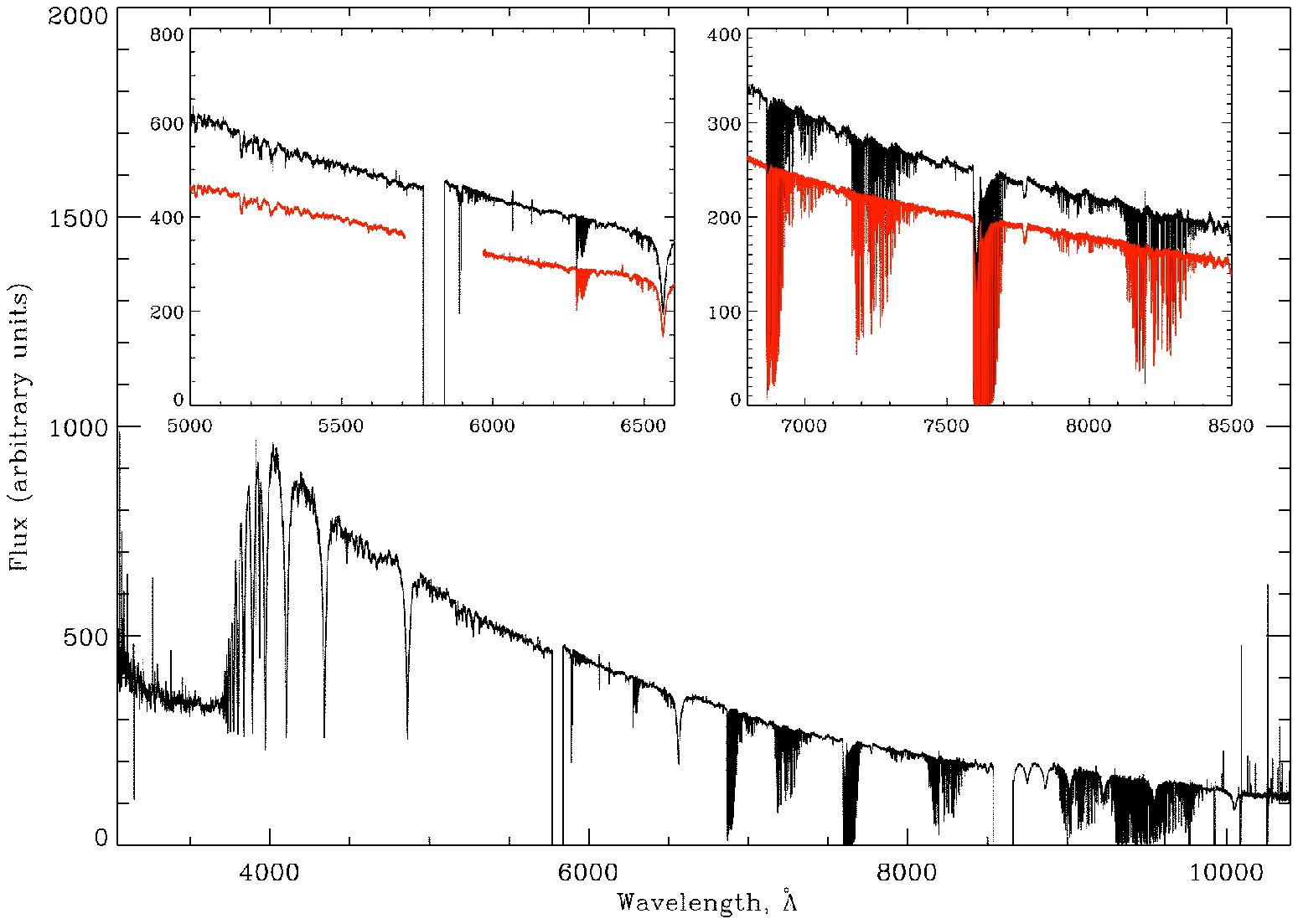}{reduced_spec}{An original spectrum of the
star HD320764 from UVES-POP.  Two zoom panels show the region around the gap
(left) and another region with significant ripples (right).  The results of
recalibration are shown in red.}

\section{UVES-POP recalibration approach}
We have re-reduced original UVES-POP data downloaded from the ESO Data
Archive using the version 5.5.7 of the UVES pipeline. We also
introduced algorithms to remove ripples and correctly merge spectral chunks. 
In order to eliminate ripples, we (i) shift the flat-field image with
respect to the science frame and (ii) add or subtract a value proportional
to the flux in order to account for imperfect scattered light correction:
these two operations are run in a loop that minimizes the flux difference
between overlapping order fragments.

We have also developed an approach to merge non-overlapping segments using
synthetic spectra (PHOENIX \citep{2013A&A...553A...6H} and BT-Settl \citep{2014IAUS..299..271A}) 
and stellar atmospheric parameters reported in the literature for UVES-POP stars: this allows us
 to estimate the flux difference between the both sides of each gap.

\section{Additional steps} 
When we merge spectral segments, we also calculate the value of extinction
E(B-V) on the line-of-sight. It allows us to deredden UVES-POP spectra for
the future use in stellar population synthesis.

In the red part of the wavelength range ($>$680~nm) it is also important to
correct the telluric absorptions originating from the Earth atmosphere. 
This will be done for the 1st data release. 
\articlefigure{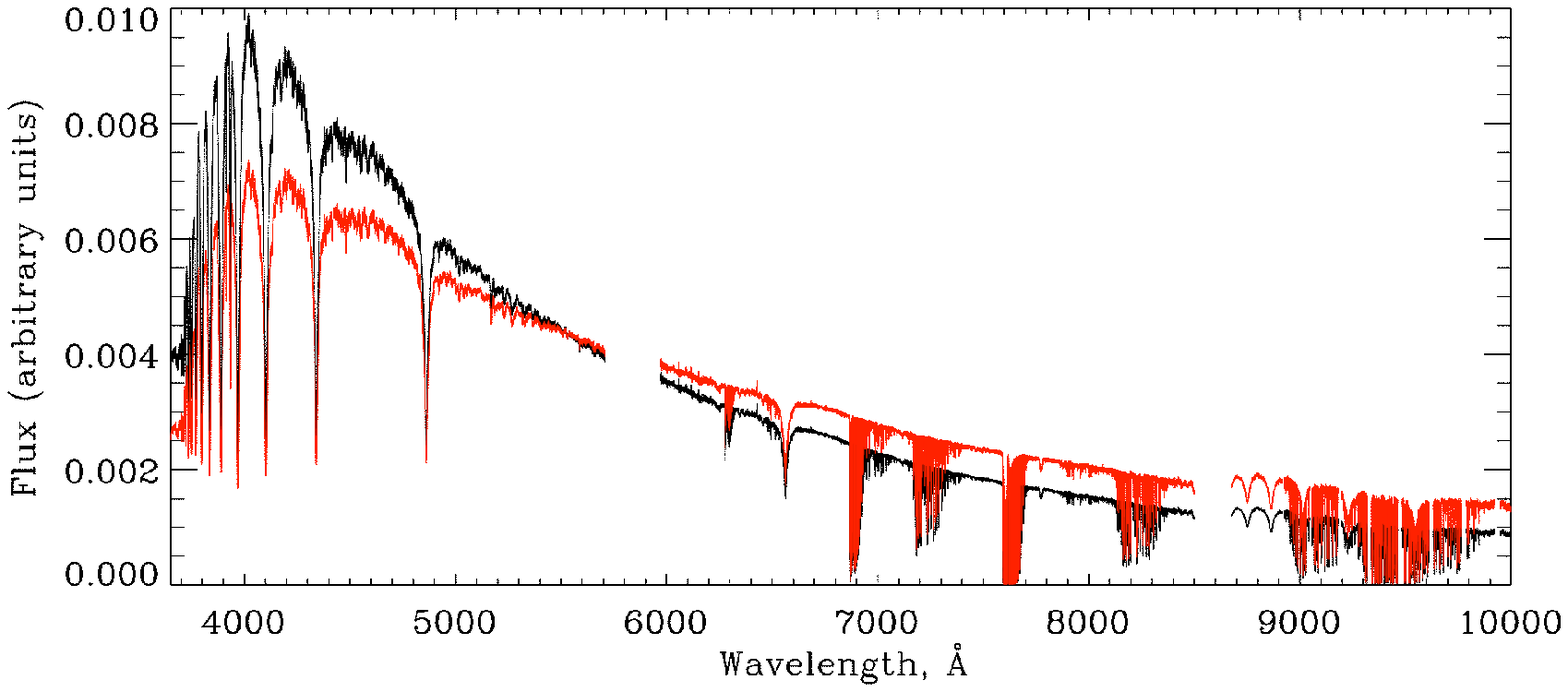}{unred}{Spectrum of the star HD320764 before (red)
and after (black) correction for the interstellar reddening.}

\section{Web-based visualisation} 
We have developed a dedicated interactive web-service to view recalibrated
UVES-POP spectra. Top left panel displays a 3D-plot of stars in the
$T_{\mathrm{eff}}$-[Fe/H]-$log(g)$ parameter space and allows one to choose
any star in order to view and download its spectrum (bottom panel]). The
preliminary version of the service is available at the following web-address:
\url{http://sl.voxastro.org/3d-viewer}

\articlefigure{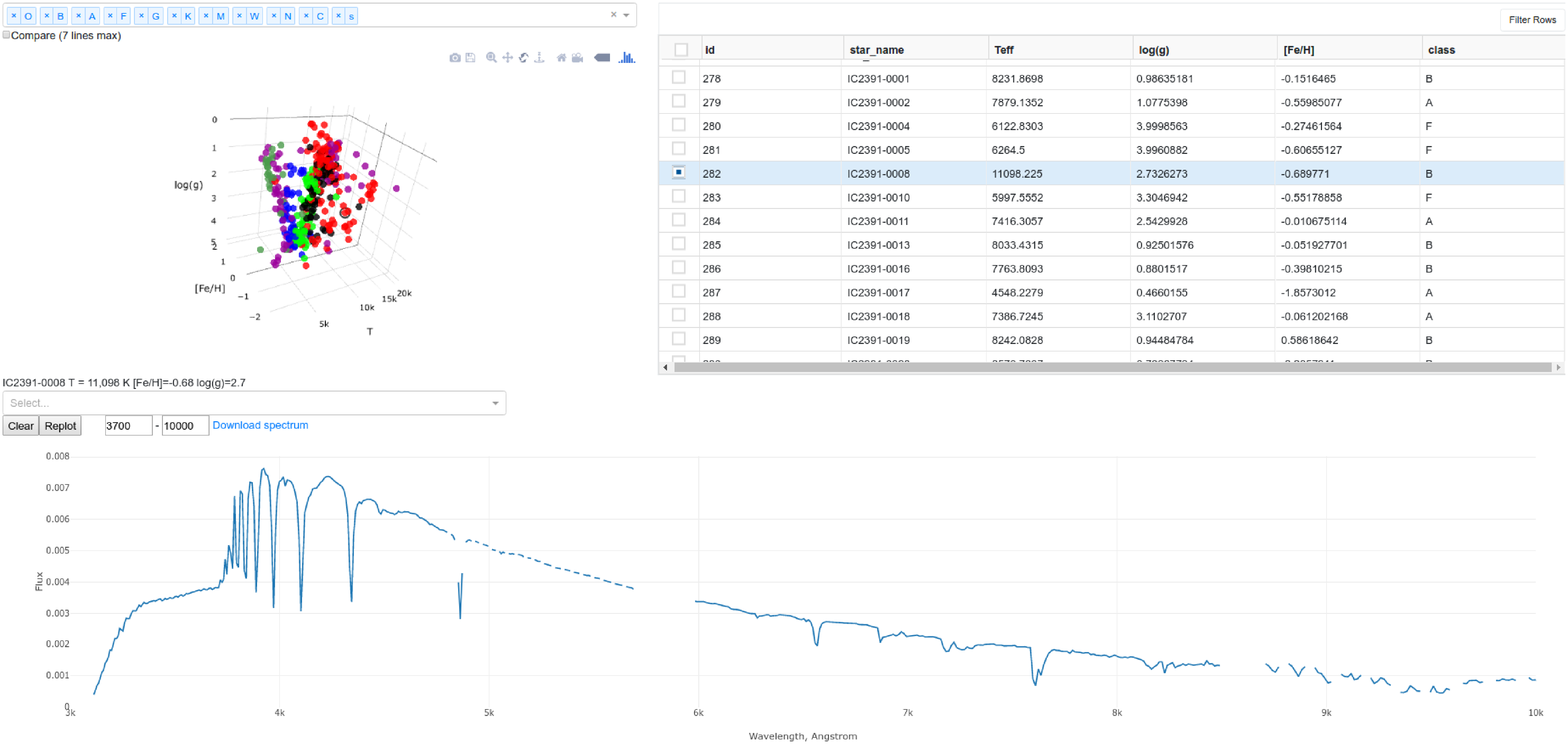}{viewer}{An interactive web-based spectral viewer
developed for the visualisation of spectra from UVES-POP and other stellar
spectral libraries.}

\acknowledgements The authors are grateful to the citizen scientist M.~Chernyshov and 
the senior researcher I.~Katkov (Moscow State University) for their valuable help with 
the development of the web-based viewer.  The authors acknowledge support from the 
RScF grant 17-72-20119.

\bibliography{P10-26}  % For BibTex

\end{document}